# Reciprocal Symmetric and Origin of Quantum Statistics

Mushfiq Ahmad

Department of Physics, Rajshahi University, Rajshahi, Bangladesh

E-mail: mushfiqahmad@ru.ac.bd

**Abstract**

Boltzmann's differential equation is replaced by the corresponding reciprocal symmetric finite difference equation. Finite difference translates discreteness of energy. Boltzmann's function, then, splits into two reciprocally related functions. One of them gives Planck's radiation relation and the other one gives the corresponding Fermi-Dirac relation.

## 1. Introduction

We have observed that spin is related to reciprocal symmetric transformation[1]. A related question is whether there is any relation between reciprocal symmetry and quantum statistics. This we shall study in this paper.

## 2. Planck's Radiation Formula and Relation to Statistics

The formula for the distribution of light in a black body is given by[2]

$$I(w)dw = <P> \frac{w^2 dw}{\pi^2 c^2} \qquad (2.1)$$

with

$$<P> = \frac{\hbar w}{\exp(\hbar w/kT) - 1} \qquad (2.2)$$

Planck's hypothesis has been invoked to find $<P>$.

The number $N_n$ of particles in nth energy level is given by Boltzmann's probability distribution relation[3]

$$N_n = N_0 \exp(-p_n.s) = \exp(-p_n.s) \qquad (2.3)$$

We have chosen $N_0 = 1$

(2.3) will correspond to (2.2) if $p_n = n\hbar w$ and $s = 1/kT$, so that

$$1/N_1 = \exp(p_1.s) = \exp(\hbar w/kT) \tag{2.4}$$

## 3. Symmetric Boltzmann Equation

$N_n$ is the solution of Boltzmann's equation[4], which corresponds to (2.3)

$$\frac{dN_n}{ds} = -p_n N_n \tag{3.1}$$

To invoke symmetry principle we replace the above differential equation by the corresponding reflection symmetric finite difference equation[5] below.

$$\frac{DN'_n}{D(s,\delta)} = -\frac{2}{\delta} PN'_n \tag{3.2}$$

where

$$\frac{DN'_n}{D(s,\delta)} = \frac{N'_n(s+\delta) - N'_n(s-\delta)}{2\delta} \tag{3.3}$$

In the limit $\delta \to 0$ (3.2) must go to (3.1)

$$\frac{DN'_n}{D(s,\delta)} = -\frac{2}{\delta} PN'_n \Rightarrow \frac{dN_n}{ds} = -p_n N_n \tag{3.4}$$

One of the solutions of (3.2) is

$$N'_n = N'_0 E(-p_n\delta/2, s/\delta) = E(-p_n\delta/2, s/\delta) \tag{3.5}$$

We have chosen $N'_0 = 1$ and

$$E(-p_n\delta/2, s/\delta) = \left(\frac{1 + -\frac{p_n\delta}{2}}{1 + \frac{p_n\delta}{2}}\right)^{s/\delta} = \exp\left(i\frac{2n\pi s}{\delta}\right) E(-p_n\delta/2, s/\delta) \tag{3.6}$$

Where n is an integer

(3.5) satisfies (3.2) with

$$P = P(p_n\delta/2) = \frac{p_n\delta/2}{1-(p_n\delta/2)^2} \tag{3.7}$$

We observe that $P(p_n\delta/2)$ is symmetric under the change $p_n\delta/2 \to -2/(p_n\delta)$ so that

$$P(-2/(p_n\delta)) = P(p_n\delta/2) \tag{3.8}$$

This means that $N''_n$ is another solution where

$$N''_n = E(2/(p_n\delta), s/\delta) \tag{3.9}$$

$N''_n$ has been obtained by replacing $p_n\delta/2 \to -2/(p_n\delta)$ in $N'_n$

$$E(2/(p_n\delta), s/\delta) = \left(\frac{1+\frac{2}{p_n\delta}}{1-\frac{2}{p_n\delta}}\right)^{s/\delta} = \left(-\frac{1+\frac{p_n\delta}{2}}{1-\frac{p_n\delta}{2}}\right)^{s/\delta}$$

$$= \exp\left(\frac{(2n+1)\pi s}{\delta}i\right) E(p_n\delta/2, s/\delta) \tag{3.10}$$

Therefore,

$$N''_n = \exp\left(i\frac{(2m+1)\pi s}{\delta}\right) E(p_n\delta/2, s/\delta) \tag{3.11}$$

In the limit as $\delta \to 0$, $N'_n$ gives the classical exponential function solution of (3.1).

$$N'_n \xrightarrow[\delta \to 0]{} N_n \exp(2n\pi/\delta i - p_n)s = \exp(-p_n s) \tag{3.12}$$

$N''_n$ is an oscillating function under the envelop

$$|N''_n| \xrightarrow[\delta \to 0]{} \exp(p_n s) = 1/N'_n \tag{3.13}$$

If we choose $m=0$, $n=1$ in (3.11) we get

$$N''_1 = -.E(p_1\delta/2, s/\delta) \tag{3.14}$$

In the limit $\delta \to 0$, we get

$$N''_1 \xrightarrow[\delta \to 0]{} -\exp(p_1/kT) = -\exp(\hbar w/kT) \tag{3.15}$$

## 4. Planck's Formula

If we use $N'_1$ instead of $N_1 = \exp(-\hbar w/kT)$ in calculating $<P>$ as in (2.2) and in the end we replace $N'_1$ by its classical limit (3.12), we get Planck's result (2.2). If we use $N''_1$ of (3.14) instead of $N'_1$ in calculating $<P>$ and we take the limiting value (3.15) we get[6]

$$<P> = \frac{\hbar w}{-\exp(-\hbar w/kT) - 1} = \frac{\hbar w'}{\exp(\hbar w'/kT) + 1} \qquad (4.1)$$

where $\hbar w' = -\hbar w$

(4.1) gives Fermi-Dirac distribution relation. The above considerations show that $N_1$ and $N'_1$ of (2.3) and (3.5) respectively are related to Bose-Einstein statistics. On the other hand $N''_1$ of (3.14) and (3.15) is related to Fermi-Dirac statistics.

**Conclusion**

Following the procedure for getting Planck' relation, we have obtained Fermi-Dirac distribution relation from reciprocal symmetry only. No postulate of quantum statistics – like exclusion principle – is invoked.

---

[1] Mushfiq Ahmad. Reciprocal Symmetry and the Origin of Spin. http://www.arxiv.org/abs/math-ph/0702043
[2] Feynman, Leighton, Sands. Feynman Lectures on Physics. Addison-Wesley Pub. Co.
[3] http://www.answers.com/topic/boltzmann-distribution
[4] Feynman, Leighton, Sands. Feynman Lectures on Physics. Addison-Wesley Pub. Co
[5] Mushfiq Ahmad. Reciprocal Symmetry and Equivalence between Relativistic and Quantum Mechanical Concepts. http://www.arxiv.org/abs/math-ph/0611024
[6] Mushfiq Ahmad. Reciprocal Symmetry and Equivalence between Relativistic and Quantum Mechanical Concepts. http://www.arxiv.org/abs/math-ph/0611024